\documentclass{article}
\usepackage{aip}
\usepackage{amsmath}
\usepackage{amsfonts}
\usepackage{amssymb}
\usepackage{graphicx}
\newtheorem{theorem}{Theorem}
\newtheorem{acknowledgement}[theorem]{Acknowledgement}

\begin{document}

\author{\textbf{A.A.Borghardt, V.M.Gokhfeld and D.Ya.Karpenko}\\\textit{Donetsk Institute of Physics and Technology}\\\textit{72 Rosa Luxemburg St., Donetsk 83114, Ukraine}\\\textit{E-mail: }\textsf{gokhfeld@host.dipt.donetsk.ua}}
\title{On the Statistical Properties of Localized Solutions of Klein-Fock-Gordon
Equation }
\date{21.10.2001 }
\maketitle
\begin{abstract}
We consider the normalized axisymmetric solutions of Klein-Fock-Gordon
equation with energy spectrum that lies below usual rest energy $mc^{2}$. It
is shown that the gas of hypothetical particles, described by such solutions,
would possess uncommon thermodynamic and kinetic characteristics, in
particular, an anomalously high temperature of Bose-Einstein condensations as
well as - in the case of charged particles - a conductivity, much exceeding
that of conventional plasma.
\end{abstract}

\subsubsection{Introduction}

As it is well known, the motion of relativistic particle of mass $m$ can be
described by Klein-Fock-Gordon (KFG) equation,
\begin{equation}
\triangle\Psi=c^{-2}\Psi_{tt}+\kappa^{2}\Psi\text{ ; }\kappa\equiv
mc/\hbar\text{ , }\label{1}%
\end{equation}
for its wave function, $\Psi.$ As a rool, just simpliest, plane wave solutions
of Eq.(1),
\begin{equation}
\Psi\propto\exp(i\mathbf{k\cdot r}-i\omega t)\text{ },\label{2}%
\end{equation}
are usually considered in empty - uniform and isotropic - space. They only
depend on energy $\hbar\omega$ and momentum $\hbar\mathbf{k}$ of a particle
and yield the standard relativistic dispersion relation,
\begin{equation}
\omega(\mathbf{k})=c\sqrt{\kappa^{2}+\left|  \mathbf{k}\right|  ^{2}}\label{3}%
\end{equation}

However, Eq.(1) also admits more general stationary solutions, not uniform in
a plane orthogonal to $\mathbf{k}$ . Basing on the correspondence principle,
we can assume that $OZ$ axis, parallel to $\mathbf{k}$ and coincident with
classical trajectory of particle, is assigned in coordinate space. So - as a
hypothesis - we can introduce into consideration the wave functions of a form
\begin{equation}
\Psi(x,y,z,t)=\Phi(\rho)\exp(ikz-i\omega t)\text{ ; }\rho\equiv\sqrt
{x^{2}+y^{2}}\text{ },\label{4}%
\end{equation}
localized nearby this axis. Axially symmetric solutions of KFG-equation and
their role in causal propagator construction have been considered in [$^{1}$].
In present communication we investigate some statistical and kinetic
properties of the ensemble of hypothetic particles described by such solutions
and demostrate that these properties differ essentially from the statistics of
conventional relativistic gas.

\subsubsection{\bigskip Dispersion relations}

Radial function $\Phi(\rho)$ satisfies the equation%

\begin{equation}
\Phi_{\rho\rho}+\rho^{-1}\Phi_{\rho}=(\kappa^{2}+k^{2}-\omega^{2}/c^{2}%
)\Phi\equiv q^{2}\Phi\text{ }.\label{5}%
\end{equation}

In addition to usual limit conditions $\Phi(+\infty)\rightarrow0$ $;$
$\Phi_{\rho}(+\infty)\rightarrow0$ , it is natural to require $\Phi$ to be
quadratically integrable in $z=const$ plane, i.e. to belong to Gilbert space
(see [$^{2}$] ). Such solutions, normalized to $1$, exist only at
\textit{positive} values of parameter $q^{2}$ and are given by McDonald
function (zeroth order modified Bessel function of the second kind):
\begin{equation}
\Phi(\rho)=(q/\pi)\mathrm{K}_{0}(q\rho)\text{ }.\label{6}%
\end{equation}

Aside from this, we have to require the stability of solution (i.e. reality of
$\omega$) at every permitted values of $k$ or, the same, to exclude
''non-physical'' states with superlight group speeds: $v\equiv\partial
\omega/\partial k\leq c$ . As a result, we come to more general, than Eq.(3),
dispersion relation:
\begin{equation}
\omega(k,q)=c\sqrt{\kappa^{2}+k^{2}-q^{2}}\text{ ; }0<q^{2}\leq\kappa
^{2}\text{ }.\label{7}%
\end{equation}

With increasing of parameter $q^{2},$ localization becomes stronger and energy
decreases. At maximal value $q=\kappa$ the wave function is localized in a
region with radius of order of Compton wavelength $\hbar/mc$ , and spectrum
begins from zero energy, like that of massless particles: $\omega
(k,\kappa)=c\left|  k\right|  $ $.$

The presence of anomalous branch of spectrum, lying under usual rest energy
$mc^{2}$ , is the result of singular eigen-functions admission. Bound
solutions of Eq.(6) can exist only at $-\infty<q^{2}\leq0$ ; they are
represented by non-normable (i.e. delocalized) Bessel functions $\mathrm{J}%
_{0}(\left|  q\right|  \rho)$ , including the plane wave mooving along $OZ$
axis at $q=0$ . However, it is necessary to indicate that such admission does
not contrary to well known principles of quantum mechanics: it is commonly
accepted (see, e.g., [$^{2}$]) that the wave function $-$ or rather $\left|
\Psi\right|  ^{2}$ $-$ may entirely have integrable singularities in domains
of dimensionalities lower thah its domain of definition. Seemingly, the most
famous example of singular wave function application is relativistic theory of
hydrogen atom [$^{3}$].

\subsubsection{\bigskip Energy density of states}

Now consider the ideal gas of N particles with dispersion law (7) in space
volume $V$, and calculate the phase volume $N(\omega)$ , i.e. quantity of
states with energies not exceeding $\hbar\omega.$ In space of wave numbers
with cylindrical coordinates $(k,q,\varphi)$ the corresponding domain is
restricted by the surface of revolution
\begin{equation}
q(k,\omega)=\sqrt{\kappa^{2}-\omega^{2}/c^{2}+k^{2}}\label{8}%
\end{equation}
and, according to inequality (7), by cylinder $q=\kappa$ (see Fig. 1). So one
can find%

\begin{equation}
N(\omega)=\frac{V}{8\pi^{3}}\int\limits_{0}^{2\pi}d\varphi\int\limits_{-\omega
/c}^{\omega/c}dk\int\limits_{\sqrt{\kappa^{2}-\omega^{2}/c^{2}+k^{2}}}%
^{\kappa}qdq=\frac{V\omega^{3}}{6\pi^{2}c^{3}}\text{ ,}\label{9}%
\end{equation}
and, by definition, the density of anomalous (localized) states with
$\hbar\omega<mc^{2}$ is
\begin{equation}
g_{L}(\omega)\equiv\partial N(\omega)/\partial\omega=V\omega^{2}/2\pi
^{2}c^{3\text{ }}\text{ }\label{10}%
\end{equation}
(see Fig. 2). Indicate that it does not depend on the mass of particle.

Eqs. (7) and (10) are the key results: the presence of activationless spectrum
and possibility of high speeds at low energies predetermine an anomalous
statistical properties of the gas of particles described by localized
solutions of KFG equation.\bigskip

\subsubsection{Bose-Einstein condensation}

We can find the degeneracy temperature of anomalous gas considering the effect
of Bose-Einstein condensation. As is known, it means that the maximal $-$ i.e.
corresponding to zero value of chemical potential $\mu$ $-$ number of excited
bosons becomes less than their given total number N at temperatures lower than
some critical one, $T_{C}$ . So the rest particles are ''condensing'' in
ground state (with minimal energy). Critical $-$ or degeneracy $-$ temperature
separates (by order of value) the ranges of application of classical and
quantum, in given case Bose statistics.

The equation for $T_{C}$ (in energy units) can be written in form
\begin{equation}
\mathrm{N=}\int\limits_{0}^{\infty}\frac{d\omega g(\omega)}{\exp(\hbar
\omega/T)-1}\text{ }\label{11}%
\end{equation}
(see, e.g.,[$^{4}$]). Since under ''earthly'' laboratory conditions we alwais
have $T<<mc^{2}$ , the expression (10) for $g(\omega)$ can be very exactly
substituted to Eq.(11) at any $\omega$ . As a result we find
\begin{equation}
\mathrm{N\cong}\frac{V}{2\pi^{2}}\left(  \frac{T}{\hbar c}\right)  ^{3}%
\int\limits_{0}^{\infty}\frac{x^{2}dx}{e^{x}-1}\equiv\frac{V}{\pi^{2}}\left(
\frac{T}{\hbar c}\right)  ^{3}\zeta(3)\text{ , }%
\end{equation}
where $\zeta(x)$ is the Riemann function, $\zeta(3)\approx$ 1.202. Critical
temperature for localized states is
\begin{equation}
T_{C}\cong\hbar c(\mathrm{N\pi}^{2}/V\zeta(3))^{1/3}\approx2\hbar
c(\mathrm{N}/V)^{1/3}\label{13}%
\end{equation}
and does not depend on mass; the share of condensed particles is
\begin{equation}
\mathrm{N}_{0}(T)/\mathrm{N}\cong1-(T/T_{C})^{3}.\label{14}%
\end{equation}

Remind that for usual Bose gas the corresponding quantities are given by
expressions
\[
(T_{C})_{cl}\approx(\mathrm{N}/V)^{2/3}\hbar^{2}/m\text{ ;}
\]
\begin{equation}
(\mathrm{N}_{0}(T)/\mathrm{N})_{cl}=1-(T/(T_{C})_{cl})^{3/2}\text{
.}\label{15}%
\end{equation}

At rational densities of particles, when $\mathrm{N}/V<<\kappa^{3}\sim
10^{31}cm^{-3}$ , the degeneracy temperature of anomalous gas (13) exceeds
much the usual one: $T_{C}/(T_{C})_{cl}\sim$ $\kappa(V/\mathrm{N})^{1/3}$ (and
the same is the ratio $mc^{2}/T_{C}$ ). E.g., even at very small densities,
$\mathrm{N}/V\sim10^{9}cm^{-3}$ one can obtain the ''room'' degeneracy
temperature $T_{C}\sim300^{\circ}K$ .

\subsubsection{Thermodynamic functions}

Using the density states found (10) and standard formulas of statistical
mechanics [$^{4,5}$] one can easily calculate all thermodynamic functions of
anomalous gas consisting of localized particles. For instance, the average
intrinsic energy related to a particle is
\begin{equation}
E(T)\cong\frac{\hbar}{\mathrm{N}}\int\limits_{0}^{\infty}\frac{d\omega\omega
g_{L}(\omega)}{\exp((\hbar\omega-\mu)/T)-1}\text{ ,}\label{16}%
\end{equation}
where chemical potential $\mu(T)$ is determined by equation of the form (11)
but with $\hbar\omega-\mu$ in exponent instead of $\hbar\omega$. At
temperatures lower than $T_{C}$ we may assume $\mu(T)\approx0$ , so similarly
to Eq.(12) we obtain
\[
E(T)\cong\frac{3\zeta(4)}{\zeta(3)}T\approx2.701\times T\text{ ;}
\]
\begin{equation}
C_{V}\approx2.701\times B\text{ ,}\label{17}%
\end{equation}
where $C_{V}$ is a specific heat; $B\approx1.380\times10^{-16}erg/^{\circ}K$
$,$ the Boltzmann constant. For the conventional gas (with usual density of
states) at $T<(T_{C})_{cl}<<mc^{2}$ one could obtain
\[
E(T)\cong\frac{3\zeta(5/2)}{2\zeta(3/2)}T\approx0.770\times T
\]
\begin{equation}
C_{V}\approx0.770\times B\text{ }\label{18}%
\end{equation}
(see [$^{4,5}$]). In case $T>>T_{C}$ we can neglect 1 in denominator of
expressions for $E$ and N, so corresponding results for anomalous gas are
\[
\mu(T)\cong T\ln\left[  \pi^{2}\frac{\mathrm{N}}{V}\left(  \frac{\hbar c}%
{T}\right)  ^{3}\right]  \cong-3T\ln\left(  \frac{T}{T_{C}}\right)  \text{ };
\]
\begin{equation}
E\cong3T\text{ ; }C_{V}\cong3B\text{ }.\label{19}%
\end{equation}

Indicate for comparison, that thermodynamic quantities of usual gas,
consisting of only delocalized particles, attain the values (19) just in
ultra-relativistic limit, when $T>>mc^{2}.$ At real temperature, if it exceeds
$T_{C}$ , the chemical potential (with substitution ($T_{C})_{cl}$ instead of
$T_{C}$ ), intrinsic energy and specific heat of usual gas are sharply half of
expressions (19) (see [$^{4}$]).

\subsubsection{\bigskip Conductivity}

Rigorously, only electrically neutral scalar bosons can be described by KFG
equation. However, in a case of charged particle every component of its spinor
wave function must satisfy the same equation. Therefore (with understandable
provisos) we have a right to consider properties of a gas of charged particles
with the dispersion law (7).

Let us assume, as it has been done in elementary theory of metal, that each of
them carries the charge $e$ , that -- on average -- is compensated by
distributed opposite charge of massive medium, almost transparent for
carriers. Using the well known results of physic kinetics$^{4}$, we can
calculate the conductivity of such system. In the limit of high frequencies
$\Omega,$ much more than reciprocal relaxation time of carriers, one can
obtain
\begin{equation}
\sigma(\Omega)=\frac{2e^{2}}{i\hbar\Omega}\int\limits_{0}^{\infty}d\omega
\frac{\partial F}{\partial\omega}\left\langle v^{2}\right\rangle \text{
,}\label{20}%
\end{equation}
where
\[
F(\omega,\mu)=\left(  1+\exp\left(  \frac{\hbar\omega-\mu}{T}\right)  \right)
^{-1}
\]
is Fermi distribution; $v\equiv\partial\omega/\partial k=c^{2}k/\omega$ $,$
the group velocity of particle; its square is averaged over the isoenergetic
surface $S(\omega)$ has been described in Section 3:
\begin{equation}
\left\langle v^{2}\right\rangle \equiv(2\pi)^{-3}\iint dS\frac{v^{2}}%
{\sqrt{v^{2}+(\partial\omega/\partial q)^{2}}}=(2\pi)^{-3}\int\limits_{-\omega
/c}^{-\omega/c}dkv^{2}\left|  \frac{q}{\partial\omega/\partial q}\right|
=\frac{\omega^{2}}{3\pi^{2}c}\text{ }\label{21}%
\end{equation}
(the specific form of dispersion law (7) has been used in these
transformations). As a result
\begin{equation}
\sigma(\Omega)=\frac{2ie^{2}T^{2}}{3\pi^{2}\hbar^{3}c\Omega}\int
\limits_{0}^{\infty}\frac{xdx}{\exp(x-\mu/T)+1}.\label{22}%
\end{equation}

The chemical potential of anomalous Fermi gas is defined, as above, from a
constancy of total number of particles:
\begin{equation}
\mathrm{N}=2\int\limits_{0}^{\infty}d\omega F(\omega,\mu)g_{L}(\omega)\text{
.}\label{23}%
\end{equation}

In degenerated limit this gives us
\begin{equation}
\mu\cong\hbar c(3\pi^{2}\mathrm{N}/V)^{1/3}\text{ , }T<<T_{C}\text{
,}\label{24}%
\end{equation}
and for non-degenerate gas ( $T>>T_{C}$ ) the expression (19) remains,
naturally, in force (with doubled volume $V$ if the spin is 1/2). Finally we
obtain for the conductivity:
\[
\sigma(\Omega)\cong\frac{i\mathrm{N}e^{2}}{m\Omega V}\cdot\kappa\left(
\frac{V}{3\pi^{2}\mathrm{N}}\right)  ^{1/3}\text{ , }T<<T_{C}\text{ };
\]
\begin{equation}
\sigma(\Omega)\cong\frac{i\mathrm{N}e^{2}}{m\Omega V}\cdot\frac{mc^{2}}%
{3T}\text{ },\text{ }T>>T_{C}\text{ }.\label{25}%
\end{equation}

The first factor in these formulas is the classical value of $\sigma(\Omega).$
Thus, in a wide region of temperature and carriers concentration values,
\begin{equation}
T,\text{ }T_{C}<<mc^{2}\text{ , }\label{26}%
\end{equation}
the high-frequency conductivity of the gas considered, as well as the square
of plasma frequency, proportional to it, would much exceed the corresponding
characteristics of conventional plasma.

\subsubsection{Conclusion}

Thus, some unusual results follow the hypothesis of ''quasi-classical''
KFG-particles described by localized wave functions, with energies below
$mc^{2}$ . In particular, Bose-Einstein condensation of a gas of such
particles would be possible at very high temperatures; in a case of plasma
containing charged localized particles, it would posess the anomalously high
conductivity and plasmons activation energy.

However, it seems that there is no evident physical principle to cut-off the
filling of localized states. Maybe, expected reduction of symmetry of wave
functions (in comparison with the plane waves) could be more probable in a
presence of external axisymmetric potential, i.e. if particles move inside
micro- or mesoscopic cavities like carbon nanotubes [$^{6}$]$.$ The
corresponding boundary problem $-$ in cylindrical scalar potential $-$ is
solved in Appenix.

\begin{acknowledgement}
\bigskip\bigskip
\end{acknowledgement}

Thanks to D.Sc. Alexander Kovalev and D.Sc. Alexander Philippov for fruitful discussion.

\bigskip

\bigskip

\subsubsection{Appendix}

\bigskip Consider the stationary axisymmetric solutions of Eq.(1) in the form
of
\begin{equation}
\Psi=\Phi(\rho)\exp(ik_{z}z-i\omega t)\text{ }\label{A1}%
\end{equation}
in cylindrical bore $\rho\equiv\sqrt{x^{2}+y^{2}}\leq R$ restricted by (for
simplicity) infinitely high potential barrier. Therefore, we assume the given
energy of particle ( $\hbar\omega\geq0$ ) and projection of its momentum onto
bore axis. Radial function $\Phi(\rho)$ must satisfy the equation
\begin{equation}
\Phi_{\rho\rho}+\rho^{-1}\Phi_{\rho}=Q^{2}\Phi\label{A2}%
\end{equation}
where
\begin{equation}
Q^{2}=\kappa^{2}+k_{z}^{2}-\omega^{2}/c^{2}\text{ ,}\label{A3}%
\end{equation}
and the boundary condition $\Phi(R)=0$ . Aside from it, by analogy with
Section 2, we require the normalization integral
\begin{equation}
M(R)\equiv2\pi\int\limits_{0}^{\infty}\rho\left|  \Phi(\rho)\right|  ^{2}%
d\rho\label{A4}%
\end{equation}
to be finite at any radius of the domain of free movement of particles,
including $R\rightarrow\infty$ . It is not difficult to certaine that the
linear combination of modified Bessel functions of first and second kind,
\begin{equation}
\Phi(\rho)\propto\mathrm{K}_{0}(Q\rho)-\frac{\mathrm{K}_{0}(QR)}%
{\mathrm{I}_{0}(QR)}\mathrm{I}_{0}(Q\rho)\text{ ,}\label{A5}%
\end{equation}
satisfies both these conditions, just at positive values of the parameter
$Q^{2}$ . Putting, as above, the additional physical condition, $v_{z}%
\equiv\partial\omega/\partial k\leq c$ , i.e. considering the sub-light
particles only, we obtain the continuous spectrum
\begin{equation}
\omega(k_{z},Q)=c\sqrt{\kappa^{2}+k_{z}^{2}-Q^{2}}\text{ ; }0<Q^{2}\leq
\kappa^{2}\text{ ,}\label{A6}%
\end{equation}
similar to Eq.(7). Its form leads to expression (10) for energy density of
localized states and so to the rest conclusions obtained above (taking into
account that in this case the expressions for conductivity in Seciton 6 are
related to an electric field directed along the canal).

Indicate once more that anomalous (inverse) spectrum (32) with the ''effective
mass'' depending on particle localization parameter, $m^{\ast}=\sqrt
{m^{2}-(\hbar Q/c)^{2}}$ , has been conditioned by quite resonable
quantum-mechanical requirement $M(R\rightarrow\infty)<\infty$ (see (30)). If,
quite the contrary, we put a more habitual condition $\left|  \Phi\right|
<\infty$ (on the analogy of classical mechanics problems), we obtain at once
the usual discret spectrum
\[
\omega(k_{z},n)=c\sqrt{\kappa^{2}+k_{z}^{2}+(\xi_{n}/R)^{2}}
\]
\begin{equation}
\text{( }\mathrm{J}_{0}(\xi_{n})=0\text{ ; }n=1,2,3,...\text{ ),}\label{A7}%
\end{equation}
parametrized by zeros $\xi_{n}$ of Bessel function and lying higher than
$mc^{2}$ .

The same consideration can easily be carried out also for a spherical
potential well. For localized states described by singular solutions of
KFG-equation within the spherical cavity, one can find the energy density
\begin{equation}
g_{L}(\omega)=\frac{V}{2\pi^{2}c^{3}}\omega\sqrt{\kappa^{2}c^{2}-\omega^{2}%
}\label{A8}%
\end{equation}
and, then, the temperature of Bose-Einstein condensation
\[
T_{C}\cong\pi\hbar\sqrt{\frac{2}{\zeta(2)}\frac{\hbar c\mathrm{N}}{mV}}\text{
,}
\]
\begin{equation}
\zeta(2)\approx1.645\text{ .}\label{A9}%
\end{equation}

If, as assumed, $\kappa\sqrt[3]{V/\mathrm{N}}>>1,$ it much exceeds the
classical value (15).

\bigskip
\end{document}